\documentclass[preprint]{jpsj2}

\def\runauthor{
Kenji \textsc{Yonemitsu}
}

\title{
Phase Transition in a One-Dimensional Extended Peierls-Hubbard Model with a Pulse of Oscillating Electric Field: I. Threshold Behavior in Ionic-to-Neutral Transition
}

\author{
Kenji \textsc{Yonemitsu}$^{1,2}$
\thanks{E-mail address: kxy@ims.ac.jp}
}

\inst{
$^1$Institute for Molecular Science, Okazaki 444-8585 \\
$^2$Graduate University for Advanced Studies, Okazaki 444-8585
}

\recdate{\today}

\abst{
Photoinduced dynamics of charge density and lattice displacements is calculated by solving the time-dependent Schr\"odinger equation for a one-dimensional extended Peierls-Hubbard model with alternating potentials for the mixed-stack organic charge-transfer complex, TTF-CA. 
A pulse of oscillating electric field is incorporated into the Peierls phase of the transfer integral. 
The frequency, the amplitude, and the duration of the pulse are varied to study the nonlinear and cooperative character of the photoinduced transition. 
When the dimerized ionic phase is photoexcited, the threshold behavior is clearly observed by plotting the final ionicity as a function of the increment of the total energy. 
Above the threshold photoexcitation, the electronic state reaches the neutral one with equidistant molecules after the electric field is turned off. 
The transition is initiated by nucleation of a metastable neutral domain, for which an electric field with frequency below the linear absorption peak is more effective than that at the peak.
When the pulse is strong and short, the charge transfer takes place on the same time scale with the disappearance of dimerization. 
As the pulse becomes weak and long, the dimerization-induced polarization is disordered to restore the inversion symmetry on average before the charge transfer takes place to bring the system neutral. 
Thus, a paraelectric ionic phase is transiently realized by a weak electric field. 
It is shown that infrared light also induces the ionic-to-neutral transition, which is characterized by the threshold behavior.
}

\kword{
photoinduced phase transition, neutral-ionic transition, TTF-CA, charge-transfer complex, time-dependent Schr\"odinger equation, dynamics
}

\begin{document}
\sloppy
\maketitle

\section{Introduction}

Photoinduced phase transitions\cite{nasu-ed} are intriguing phenomena, where relatively weak perturbation causes a macroscopic change in magnetic, dielectric, optical and structural properties. 
It would be important to study their mechanisms to explore possibilities of dynamically controlling these properties by making use of the cooperativity that often allows spontaneous symmetry breaking. There are also possibilities of separately controlling some of these electronic properties on the respective time scales. 

Now many materials are known to show photoinduced phase transitions. 
The characters of the transitions are very different, depending on the relative importance of the interaction responsible for the cooperativity to couplings with environmental surroundings. 
In the spin crossover complexes \cite{Ogawa_PRL,Tayagaki_PRL,Enachescu_JPCB}, the spin state changes very slowly compared with molecular vibrations coupled with it. The initial and final spin states have their respective diabatic potentials as a function of the so-called interaction mode, whose crossing gives the barrier height to be overcome by thermal lattice fluctuations. The details of lattice dynamics is unimportant, so that we can safely integrate over the thermally fluctuating lattice displacements \cite{nagaosa89}. The evolution of the spin state is thus described in a stochastic manner, e.g., by using a master equation \cite{Boukheddaden_EPJB,Hoo_EPJB} or Monte Carlo simulations \cite{Romstedt_JPCS,Sakai_JPSJ}. In these materials, the direct interaction among rather distant $ d $-electron spins is much weaker than the effective interaction, which is mediated by the surrounding lattice displacements. 

In contrast, in the mixed-stack organic charge-transfer complex, TTF-CA (TTF=tetrathiafulvalene, CA=chloranil) \cite{koshihara90,koshihara95,koshihara99,suzuki99,iwai02,luty02,collet03,guerin04}, whose model is studied in this paper, the dynamics of charge density and that of dimerization-induced, staggered lattice displacements are strongly coupled \cite{miyashita03}. They are governed by the electron-lattice and electron-electron interactions \cite{huai00} that are responsible for the electronic and structural properties in equilibrium. The electronic state substantially evolves according to these interactions before the energy supplied by the photoexcitation significantly dissipates into unimportant degrees of freedom such as intramolecular vibrations. In other words, on the time scale of observation, which is rather short for TTF-CA, memories are not easily lost as in the slowly evolving spin crossover complexes. The rapidness of the electronic evolution is generally expected for a low energy barrier, i.e., when the hysteresis is small or when the state is near the instability within the hysteresis. 

Indeed the coherent motion of the macroscopic neutral-ionic domain boundary is observed \cite{iwai02}, which would disappear if the energy dissipation were significant. The evolution of the electronic state should thus be described in a deterministic manner, e.g., by combining the time-dependent Schr\"odinger equation for the electronic wave function and the classical equation of motion for the lattice displacements \cite{miyashita03}. We have actually described in ref.~\citen{miyashita03} the charge-lattice dynamics that does not follow the adiabatic potential that is projected on the plane spanned by the shape and size of a metastable domain \cite{huai00}. In this paper, we again use the Hartree-Fock (HF) approximation and introduce photoexcitations not by simply changing the occupancy of the HF orbitals \cite{miyashita03} but by incorporating a pulse of oscillating electric field into the Peierls phase of the transfer integral. Thus, the frequency, the amplitude, and the duration of the pulse can be independently varied. 

Concerning the issue of whether the evolution is stochastic or deterministic, we note that the photoinduced charge-density-wave(CDW)-to-charge-polarization(CP) phase transition in one-dimensional halogen-bridged binuclear platinum complexes show both the stochastic and deterministic characters \cite{yonemitsu03}. The dependence of the threshold photoexcitation on the electron-lattice coupling is intuitively explained by the dependence of the barrier height \cite{yonemitsu03}. Transition processes between the two electronic states may be described in a stochastic manner with the help of their diabatic potentials if only this fact is to be explained. However, the difficulty of the CP-to-CDW phase transition compared with the CDW-to-CP one needs detail explanations based on charge-transfer processes in the respective electronic states \cite{yonemitsu03}. These processes are easily missed if one uses a stochastic approach.

Among the materials that show rather rapid photoinduced charge-lattice-coupled dynamics, the TTF-CA complex would be the most extensively studied \cite{koshihara90,koshihara95,koshihara99,suzuki99,iwai02,luty02,collet03,guerin04}, which allows comparisons with model calculations. In this paper, we discuss the transition from the ionic phase to the neutral one by intra-chain charge-transfer photoexcitations, for which the coherent motion of the macroscopic neutral-ionic domain boundary is observed as well as the threshold photoexcitation density \cite{iwai02}. In the case of intramolecular excitations, a new ionic phase with disordered polarizations is suggested to appear by comparing the photoreflectance and the second-harmonic-generation (SHG) signal \cite{luty02} and by the x-ray diffraction \cite{guerin04}. The SHG intensity, which corresponds to the degree of the inversion-symmetry breaking, decreases after the photoirradiation much faster than the ionicity estimated from the photoreflectance. The x-ray diffraction gives more direct evidence, but the error bar is not so small at present.

In this paper, we extend the previous study \cite{miyashita03}, where the dynamics of charge density and that of lattice displacements during the photoinduced ionic-to-neutral phase transition is discussed in a model for TTF-CA. Here we directly incorporate a pulse of oscillating electric field. Thus, we can now study the dependence on the frequency, the amplitude, and the duration of the pulse. Furthermore, this approach allows us to discuss a possibility for an infrared-light-induced transition without producing excitons, as shown in the present paper. It also allows us to discuss the effect of a double pulse with different intervals, as demonstrated later in paper III.

\section{Extended Peierls-Hubbard Model with Alternating Potentials}\label{model}

The highest occupied molecular orbital (HOMO) at the donor site and the lowest unoccupied molecular orbital (LUMO) at the acceptor site are necessary to describe the electronic state. 
In the limit of vanishing transfer integral $t_0 \to 0$, the HOMO is doubly occupied in the neutral phase, while both of the HOMO and the LUMO are singly occupied in the ionic phase.
Different types of electron-lattice couplings have so far been employed to explain the dimerization \cite{nagaosa86-3,painelli02,kxy02,sakano96}.
Here we assume that the Coulomb interaction strength is modified by the lattice displacement \cite{sakano96}.
We use a one-dimensional extended Peierls-Hubbard model with alternating potentials at half filling, which is a single-chain analogue of the model used in ref.~\citen{huai00},
\begin{equation}
H =  H_\mathrm{el} + H_\mathrm{lat} \;,\label{g-ham} 
\end{equation}
with
\begin{align}
H_\mathrm{el} = & 
-t_0 \sum _{\sigma,l=1}^{N}
   \left( c^{\dagger }_{l,\sigma}c_{l+1,\sigma}
     + \mathrm{h.c.} \right)  \nonumber \\ 
& +\sum _{l=1}^{N} 
\left[ U n_{l,\uparrow} n_{l,\downarrow} 
+ (-1)^{l} \frac{d}{2} n_{l} \right] \nonumber \\ 
& +\sum _{l:odd}^{N} 
\bar{V}_{l} (n_{l}-2) n_{l+1}
+ \sum _{l:even}^{N} 
\bar{V}_{l} n_{l} (n_{l+1}-2) \;, \label{e-ham} \\
H_\mathrm{lat} =& \sum _{l=1}^{N}
\left[ \frac{k_1}{2}y_{l}^{2}
+\frac{k_2}{4}y_{l}^{4}
+\frac{1}{2}m_{l}\dot{u}_{l}^{2} \right] \;,\label{l-ham} 
\end{align}
where, $ c^{\dagger }_{l,\sigma} $  ($ c_{l,\sigma} $) is the creation (annihilation) operator of a $\pi$-electron with spin $\sigma$ at site $l$, $ n_{l,\sigma} = c^{\dagger}_{l,\sigma} c_{l,\sigma} $, $ n_{l} = n_{l,\uparrow} + n_{l,\downarrow} $, $ u_{l} $ is the dimensionless lattice displacement of the $l$th molecule along the chain from its equidistant position, and $ y_{l} = u_{l+1} - u_{l} $.
The distance between the $l$th and $(l+1)$th molecules is then given by $r_l=r_0 (1+u_{l+1}-u_l)$, where $r_0$ is the averaged distance between the neighboring molecules along the chain.
The donor and acceptor molecules are located at the odd and even sites, respectively.
The nearest-neighbor repulsion strength between the $ l $th and $ (l+1) $th sites $ \bar{V}_{l} $ depends on the bond length $ y_{l} $, $ \bar{V}_{l} = V + \beta _{2} y^{2}_{l} $, where $ V $ is for the regular lattice, and $\beta _{2}$ is the quadratic coefficient. 
The linear coefficient is neglected for simplicity.
The parameter $ t_0 $ denotes the nearest-neighbor transfer integral, $ U $ the on-site repulsion strength, and $ d $ the level difference between the HOMO and the LUMO in the neutral phase with $t_0=0$. 
The elastic energy is expanded up to the fourth order: the parameters $ k_{1} $ and $ k_{2} $ are the linear and nonlinear elastic constants. 
The mass of the $ l $th molecule is denoted by $ m_l $. 
The number of sites is denoted by $N$.

In the unrestricted Hartree-Fock (HF) approximation, we iteratively solve the eigenvalue equation to obtain the ground state that is self-consistent with the one-body densities $\langle c^\dagger_{l,\sigma}c_{l,\sigma} \rangle$ and $\langle c^\dagger_{l,\sigma} c_{l+1,\sigma} \rangle$ and the lattice displacements $y_l$, which satisfy the Hellmann-Feynman theorem. We then add random numbers to the initial $ y_{l} $ and $ \dot{u}_l $  values according to the Boltzmann distribution at a fictitious temperature $\it{T}$. With thus modified lattice displacements, the HF Hamiltonian is diagonalized once again by imposing the self-consistency on the one-body densities only.

Photoexcitations are introduced by modifying the transfer integral in the kinetic part of the electronic Hamiltonian, 
\begin{equation}
- \sum _{\sigma,l=1}^{N}
   \left( t_0(t) c^{\dagger }_{l,\sigma}c_{l+1,\sigma}
     + \mathrm{h.c.} \right)
\;,
\end{equation}
with the Peierls phase, 
\begin{equation}
t_0(t) = t_0 \exp[{\rm i}\frac{e r_0}{\hbar c} A(t)]
\;,
\end{equation}
where $ e $ is the absolute value of the electronic charge, and $ c $ the light velocity. 
The time-dependent vector potential $ A(t) $ is related to the electric field $ E(t) $ by 
\begin{equation}
A(t) = -c \int^{t} {\rm d}t' E(t')
\;,
\end{equation}
where we consider a pulse field, 
\begin{equation}
E(t) = E_{\rm ext} \sin \omega_{\rm ext} t
\;,
\end{equation}
with amplitude $ E_{\rm ext} $ and frequency $ \omega_{\rm ext} $ for $ 0 < t < N_{\rm ext} T_{\rm ext} $ with integer $ N_{\rm ext} $. $ E(t) $ is zero otherwise. The period is denoted by $ T_{\rm ext} $, which is given by $ T_{\rm ext} = 2 \pi / \omega_{\rm ext} $.

Now we mention how the collective nature is incorporated into this photoexcitation. 
The excitonic effect, namely, the attraction between an excited electron and a hole to form a bound pair can be included, e.g., in the random phase approximation (RPA).
The RPA treats all the linear quantum fluctuations around the unrestricted HF solution \cite{kxy93}. These fluctuations correspond to the infinitesimal deviations from the unrestricted HF state, all of which as well as their interactions are included in the time-dependent HF theory.
Deviations from the unrestricted HF state are no longer infinitesimal when we solve the time-dependent Schr\"odinger equation after the photoexcitation. Thus, the excitations generally have a finite lifetime in the present numerical calculations.
The present treatment of photoexcitation does include the excitonic effect from the beginning, as clearly shown later in the linear absorption spectrum.

For the electronic part, we thus solve the time-dependent Schr\"odinger equation, as explained in ref.~\citen{miyashita03}, during and after the photoexcitation. 
The exponential operator is decomposed \cite{suzumasu93}, in such a way that the decomposition is accurate to the order of $(\Delta t )^2 $ \cite{ono90}. 
For the lattice part, we solve the classical equation of motion, as explained in ref.~\citen{miyashita03}. The leapfrog method is employed, which is accurate to the order of $(\Delta t)^2$. 
After the electric field is turned off, the total energy is conserved within the numerical accuracy.

\section{Results and Discussions}\label{results}
	
We use $N$=100, $ t_0 $=0.17eV, $ U $=1.528eV, $ V $=0.604eV, $ d $=2.716eV, $ \beta_2 $=8.54eV, $ k_1 $=4.86eV, $ k_2 $=3400eV, and the bare phonon energy $ \omega_{\rm opt} \equiv (1/r_0)(2 k_1/m_r)^{1/2} $=0.0192eV, and impose the periodic boundary condition. 
The reduced mass $ m_r $ is defined as $ m_r = m_\mathrm{D} m_\mathrm{A} / ( m_\mathrm{D} + m_\mathrm{A} ) $ with $ m_\mathrm{D} $ for the donor molecule and $m_\mathrm{A}$ for the acceptor molecule. 
With these parameters, the dimerized ionic phase is stable and the neutral phase is metastable.
These parameters are the same as in ref.~\citen{miyashita03}, so that the bare phonon energy $ \omega_{\rm opt} $ used here is about five times higher than the optical phonon energy of the TTF-CA complex. The results are very similar for smaller $ \omega_{\rm opt} $ values including the one for the TTF-CA complex \cite{miyashita_D}, so that the conclusion is not altered by specific values of $ \omega_{\rm opt} $. Simply because it reduces the computational burden, we adopt this value. 
The ionicity is defined as $ \rho = 1 + (1/N)  \sum_{l=1}^{N} (-1)^{l} \langle n_{l} \rangle $. 
The staggered lattice displacement is defined as $ y_{st} = (1/N)  \sum_{l=1}^{N} (-1)^l y_l $.

\subsection{Linear absorption}\label{subsec:linear_absorption}

\begin{figure}
\includegraphics[height=6cm]{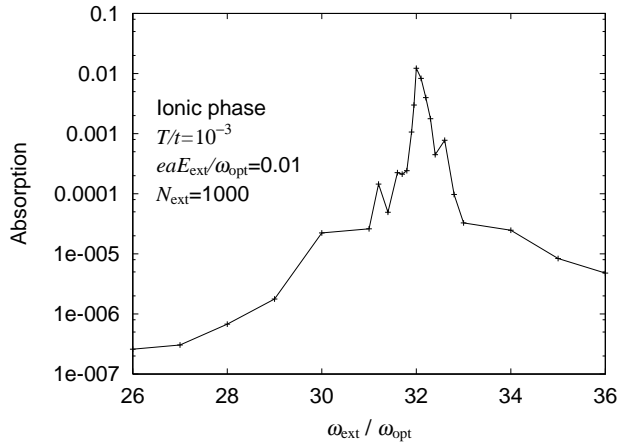}
\caption{Linear absorption, as a function of $ \omega_{\rm ext} / \omega_{\rm opt} $ in the excitonic range, in the ionic phase with lattice fluctuations at $ T/t $=10$^{-3}$. The applied electric field is very weak, $ eaE_{\rm ext}/\omega_{\rm opt} $=0.01, and lasts long, $ N_{\rm ext} $=1000.}
\label{fig:linear_absorption}
\end{figure}

First, we turn on a very weak electric field for a long time and observe the increment of the total energy. It is plotted in Fig.~\ref{fig:linear_absorption}, as a function of the frequency of the electric field. Since we here use $ \omega_{\rm opt} $=0.0192eV, the position of the absorption peak is $ \omega_{\rm ext} \simeq 32 \omega_{\rm opt} \simeq $0.61eV, which corresponds to the peak in the imaginary part of the current-current correlation function obtained by the RPA \cite{miyashita03}. The calculation of the linear response in this subsection is easily extended to those of nonlinear responses including the electromodulation spectrum \cite{kuwabara_unpublished}. Initial lattice fluctuations at $ T/t $=10$^{-3}$, i.e., $ T/\omega_{\rm opt} \simeq $0.0089 broaden the absorption spectrum. The spectrum is smooth especially when we apply a very weak field for a very long time, which would effectively average random distributions of the lattice displacements. It is reminiscent of the ergodic theorem that guarantees the replacement of the distribution average by the time average. Since the ionic phase is directly affected by the nearest-neighbor repulsion $ \bar{V}_{l} $ and thus by the lattice displacements, its absorption spectrum is much more sensitive to lattice fluctuations than that in the neutral phase in the present model.

\subsection{Threshold behavior}\label{subsec:threshold_behavior}

\begin{figure}
\includegraphics[height=12cm]{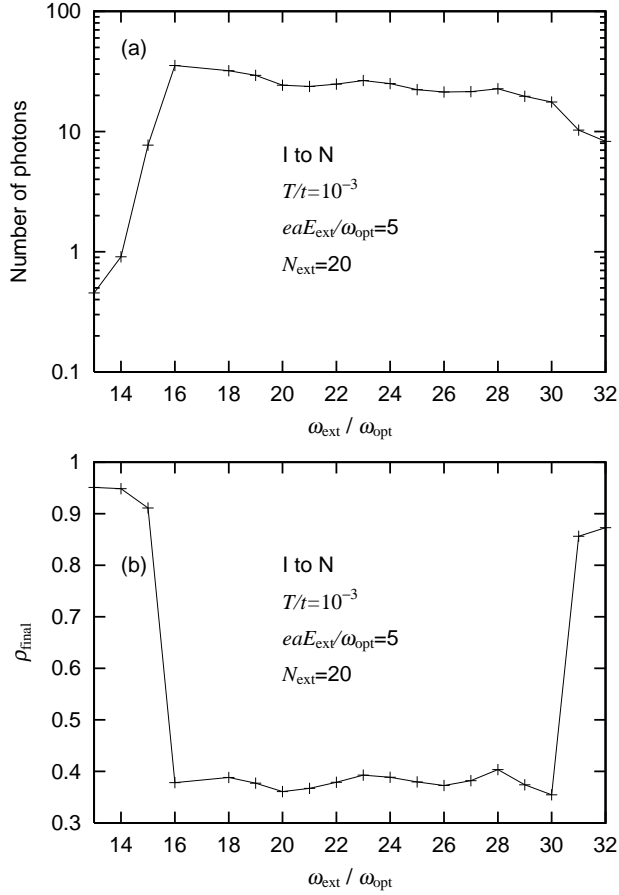}
\caption{(a) Number of absorbed photons and (b) final ionicity, as a function of $ \omega_{\rm ext} / \omega_{\rm opt} $ in the excitonic range, when the electric field with $ eaE_{\rm ext}/\omega_{\rm opt} $=5 and $ N_{\rm ext} $=20 is applied to the ionic phase at $ T/t $=10$^{-3}$.}
\label{fig:nonlinear_absorption}
\end{figure}

To see how the linear and nonlinear absorption spectra are different, we now use a stronger and shorter pulse. By adopting $ eaE_{\rm ext}/\omega_{\rm opt} $=5 and $ N_{\rm ext} $=20, we again calculate the increment of the total energy. In Fig.~\ref{fig:nonlinear_absorption}, the increment is divided by the frequency of the electric field, which corresponds to the number of absorbed photons in the 100-site chain. The strong absorption is extended to rather low frequencies [Fig.~\ref{fig:nonlinear_absorption}(a)]. In the frequency range where many photons are absorbed, the system drastically changes, namely, the transition takes place to the neutral phase. Accordingly, the ionicity finally drops [Fig.~\ref{fig:nonlinear_absorption}(b)].

\begin{figure}
\includegraphics[height=12cm]{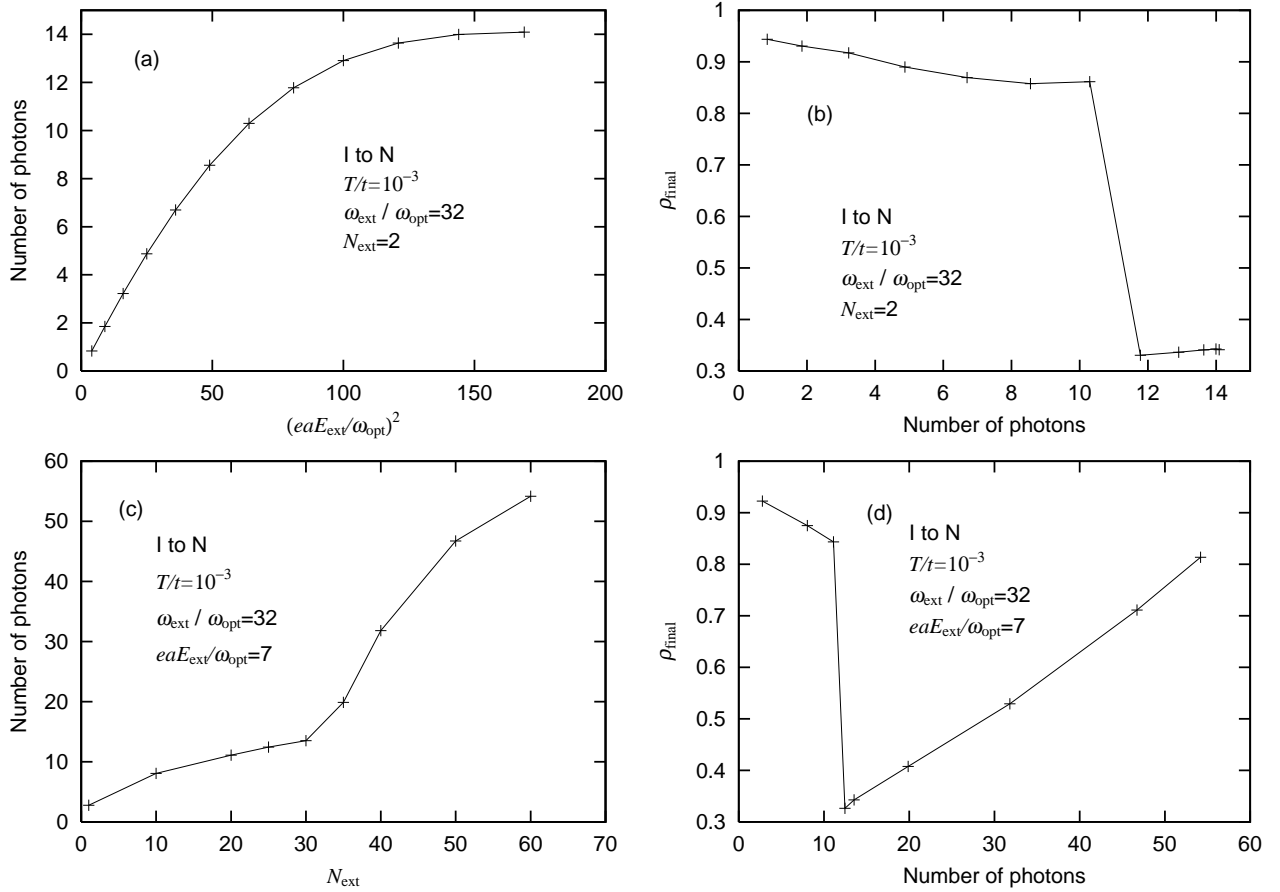}
\caption{(a) Number of absorbed photons as a function of the field strength $ ( eaE_{\rm ext}/\omega_{\rm opt} )^2 $, and (b) the corresponding final ionicity, for $ N_{\rm ext} $=2. (c) Number of absorbed photons as a function of the pulse duration $ N_{\rm ext} $, and (d) the corresponding final ionicity, for $ eaE_{\rm ext}/\omega_{\rm opt} $=7. The electric field of frequency $ \omega_{\rm ext} / \omega_{\rm opt} $=32 near the linear-absorption peak is applied to the ionic phase at $ T/t $=10$^{-3}$.}
\label{fig:pulse_dependence_on_resonance}
\end{figure}

By setting the frequency of the electric field $ \omega_{\rm ext} $ near the linear-absorption peak, $ \omega_{\rm ext}/\omega_{\rm opt} = 32 $, we vary the field strength $ ( eaE_{\rm ext} )^2 $ and the pulse duration $ N_{\rm ext} T_{\rm ext} $ in Fig.~\ref{fig:pulse_dependence_on_resonance} to see how they change the number of absorbed photons and the final ionicity. When the pulse is weak or short, the number of absorbed photons is almost proportional to the field strength or the pulse duration. This situation is smoothly connected to the linear absorption shown in Fig.~\ref{fig:linear_absorption} (i.e., where the field is infinitesimally small). When the number of photons is more than about eleven, the absorption strongly deviates from the linearity. With stronger fields, the absorption tends to be saturated [Fig.~\ref{fig:pulse_dependence_on_resonance}(a)].  Then, the final ionicity quickly drops, demonstrating that the transition to the neutral phase takes place above the threshold absorption of about eleven photons in the 100-site chain [Fig.~\ref{fig:pulse_dependence_on_resonance}(b)]. In the previous study \cite{miyashita03}, photoexcitations were introduced by simply exchanging the occupancy of the initial HF orbitals around the Fermi level. With the same random numbers added to the lattice variables (i.e., in the same initial condition), the threshold absorption was about four photons in the 100-site chain. The more realistic, oscillating electric field is therefore less effective than the simple exchange of the orbital occupancy. The absorption increases with the pulse duration more rapidly when the number of absorbed photons is more than about eleven [Fig.~\ref{fig:pulse_dependence_on_resonance}(c)]. The threshold absorption is again given by about eleven photons in the 100-site chain. Above the threshold pulse duration, the final ionicity gradually increases [Fig.~\ref{fig:pulse_dependence_on_resonance}(d)]. This indicates that, after the transition to the neutral phase is completed, further application of the oscillating electric field leads to further absorption of photons [Fig.~\ref{fig:pulse_dependence_on_resonance}(c)], which results in random distribution of electrons in the present calculations without dissipation. Because the orbitals are delocalized, the completely random distribution of electrons in the thermodynamic limit gives one electron per site, corresponding to $ \rho $=1, at half filling. In all cases, such a disordered state with $ \rho \simeq $1 is always reached by sufficiently long application of the oscillating electric field.

\begin{figure}
\includegraphics[height=12cm]{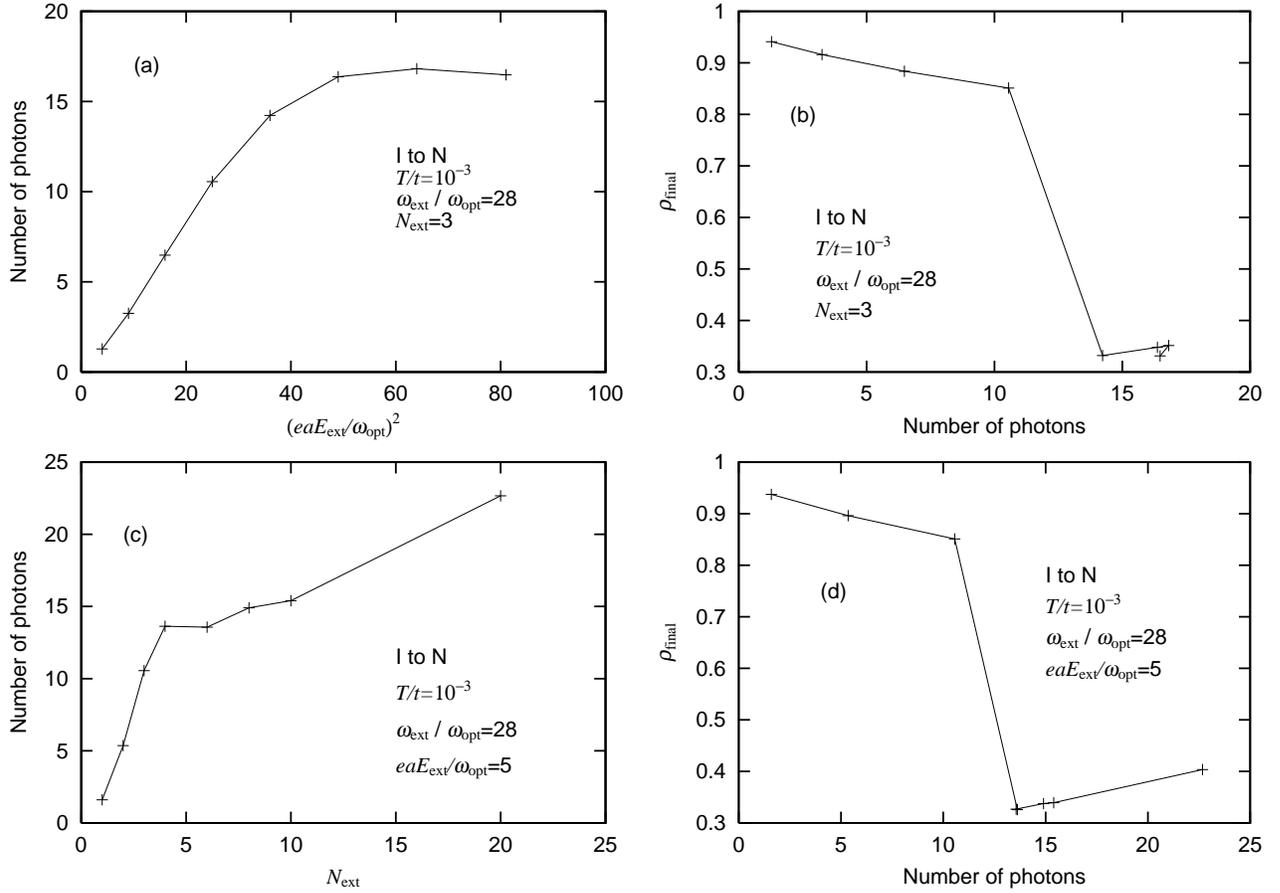}
\caption{(a) Number of absorbed photons as a function of the field strength $ ( eaE_{\rm ext}/\omega_{\rm opt} )^2 $, and (b) the corresponding final ionicity, for $ N_{\rm ext} $=3. (c) Number of absorbed photons as a function of the pulse duration $ N_{\rm ext} $, and (d) the corresponding final ionicity, for $ eaE_{\rm ext}/\omega_{\rm opt} $=5. The electric field of frequency $ \omega_{\rm ext} / \omega_{\rm opt} $=28 below the linear-absorption peak is applied to the ionic phase at $ T/t $=10$^{-3}$.}
\label{fig:pulse_dependence_off_resonance}
\end{figure}

Now we use a frequency $ \omega_{\rm ext} $ below the linear-absorption peak, $ \omega_{\rm ext}/\omega_{\rm opt} = 28 $, in Fig.~\ref{fig:pulse_dependence_off_resonance} to see how the above behavior is modified when the field is off resonance. It is generally similar to that when the field is on resonance. Namely, when the number of photons is more than about eleven, the absorption strongly deviates from the linearity [Figs.~\ref{fig:pulse_dependence_off_resonance}(a) and \ref{fig:pulse_dependence_off_resonance}(c)], and the transition to the neutral phase takes place above this threshold absorption [Figs.~\ref{fig:pulse_dependence_off_resonance}(b) and \ref{fig:pulse_dependence_off_resonance}(d)]. However, to go beyond the threshold absorption, the pulse duration with $ N_{\rm ext} $=4 is sufficient for $ eaE_{\rm ext}/\omega_{\rm opt} $=5 [Figs.~\ref{fig:pulse_dependence_off_resonance}(c) and \ref{fig:pulse_dependence_off_resonance}(d)], which is compared with $ N_{\rm ext} $=25 for $ eaE_{\rm ext}/\omega_{\rm opt} $=7 [Figs.~\ref{fig:pulse_dependence_on_resonance}(c) and \ref{fig:pulse_dependence_on_resonance}(d)] on resonance. The frequency $ \omega_{\rm ext} / \omega_{\rm opt} $=28 is located in the low-frequency tail of the linear absorption. The efficiency in the off-resonant condition comes from the fact that excitons localized due to lattice fluctuations nucleate neutral domains more easily than delocalized excitons. This fact will be discussed later again.

At the threshold absorption, we estimate the ratio of the energy needed for the phase transition to the energy supplied by the photoirradiation. The energy difference between the stable ionic phase and the metastable neutral phase is 0.16eV for the 100-site chain (Fig.~1 of ref.~\citen{miyashita03}). The energy to create two neutral-ionic domain walls, which are accompanied with each metastable domain, is about 0.3eV according to Fig.~10 of ref.~\citen{huai00}. (It is about a half of that in the strong-coupling limit, $ V $=0.604eV \cite{nagaosa86-2}.) Meanwhile, the frequency of the oscillating electric field used here is either $ 32 \omega_{\rm opt} \simeq $0.61eV, which corresponds to the energy of a delocalized exciton at the linear absorption peak, or $ 28 \omega_{\rm opt} \simeq $0.54eV. Therefore, a single photon is energetically sufficient to bring about the transition. The increment of the total energy due to initially given lattice fluctuations is 0.07eV before the photoirradiation, so that it is much smaller than the values above. Just above the threshold absorption, the lattice kinetic energy is finally in the range between 0.1eV and 0.2eV for the 100-site chains, and the lattice elastic energy is comparable with it, in all the cases shown in Figs.~\ref{fig:pulse_dependence_on_resonance}, \ref{fig:pulse_dependence_off_resonance} and \ref{fig:strength_dependent_threshold}.

These values indicate that more than 90\% of the energy supplied by the photoirradiation is absorbed into the electronic state, leading to deviation from the static self-consistent solution in the metastable neutral phase. The final ionicity apparently deviates from 0.2 in the static solution, as shown in Fig.~10 of ref.~\citen{miyashita03}. Thus, the threshold density of absorbed photons (eleven photons in the present case, four photons by the simple exchange of the orbital occupancy \cite{miyashita03}, in the 100-site chain) is not important. Important is the fact that less than 10\% (about 20\% if the orbital occupancy is exchanged) of the supplied energy is used to bring about the transition along the fact that this ratio is insensitive to whether the field is on resonance or off resonance. Later, we will show that the ratio of the energy difference between the two static electronic states to the energy actually supplied by the photoirradiation somewhat depends on the field strength or the pulse duration. The ratio becomes low as the field is weakened. In general, as the energy dissipates more into processes other than charge transfers, the ratio becomes low. To theoretically estimate the ratio in this many-body system is far beyond the scope of the present study. It will need accumulations of numerical calculations.

\begin{figure}
\includegraphics[height=12cm]{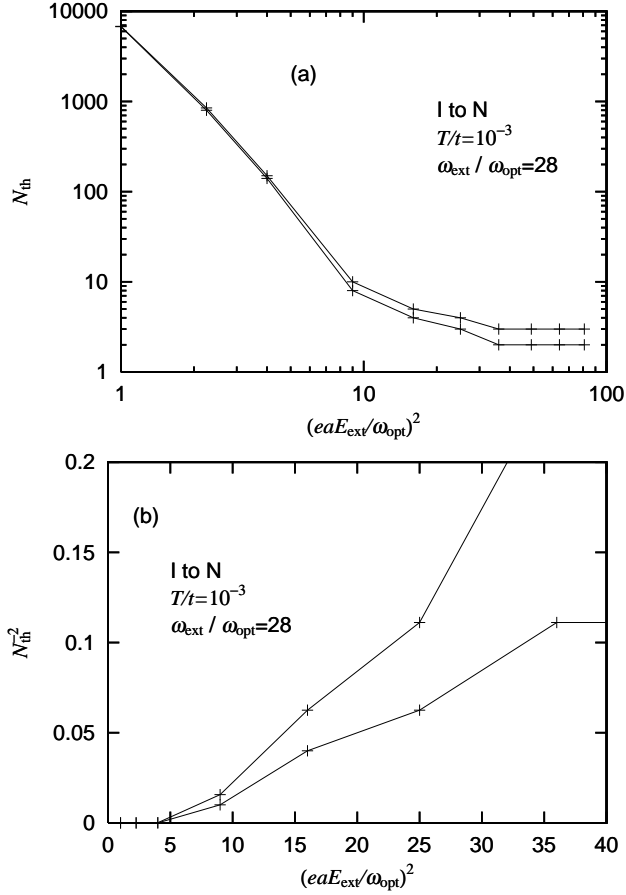}
\caption{(a) Pulse duration $ N_{\rm ext} $ just above and below the threshold $ N_{\rm th} $, and (b) the inverse of its square ($ \sim N_{\rm th}^{-2} $), as a function of the field strength $ ( eaE_{\rm ext}/\omega_{\rm opt} )^2 $. The electric field of frequency $ \omega_{\rm ext} / \omega_{\rm opt} $=28 below the linear-absorption peak is applied to the ionic phase at $ T/t $=10$^{-3}$.}
\label{fig:threshold}
\end{figure}

We have calculated the evolution of the ionicity with various strengths and durations of pulses to find their relation at the threshold. The pulse duration $ N_{\rm ext} $ just above and below the threshold for the transition is plotted in Fig.~\ref{fig:threshold}(a), as a function of the field strength $ ( eaE_{\rm ext}/\omega_{\rm opt} )^2 $. In general, to describe the dynamics of the density of excited states under photoirradiation, a master equation is useful when thermal transitions are so dominant that the evolution is described in a stochastic manner \cite{nagaosa89}. In such a case, it is known that the time required for a transition $ t_{\rm cross} $ is proportional to $ ( I - I_{\rm th} )^{-1/2} $ when the field strength $ I $ is just above the threshold $ I_{\rm th} $ \cite{nagaosa89}. Then, we also plot $ N_{\rm ext}^{-2} $ in Fig.~\ref{fig:threshold}(b), as a function of the strength.  If the same law held as in the stochastic case, $ N_{\rm ext}^{-2} $ would be a linear function of $ ( eaE_{\rm ext}/\omega_{\rm opt} )^2 $ and would vanish at a finite value of $ ( eaE_{\rm ext}/\omega_{\rm opt} )^2 $, for which the transition would never take place no matter how long the field is applied. However, the present result shows a nonlinear curve, suggesting the absence of $ I_{\rm th} $ although it is very hard to demonstrate it numerically. Even if the field is very weak, e.g., $ ( eaE_{\rm ext}/\omega_{\rm opt} )^2 $=1, the transition finally takes place after the field is applied for a very long time. Thus, in the deterministic case, the conversion occurs more efficiently than in the stochastic case. In the experimental condition, if a weak field is applied, a portion of energy would flow into the environment (i.e., outside the model system) before the transition occurs, so that the efficiency would become lower. As the exchange of energy between the system and the environment dominates the direct exchange inside the system, the efficiency would approach the value predicted in the stochastic case.

\begin{figure}
\includegraphics[height=12cm]{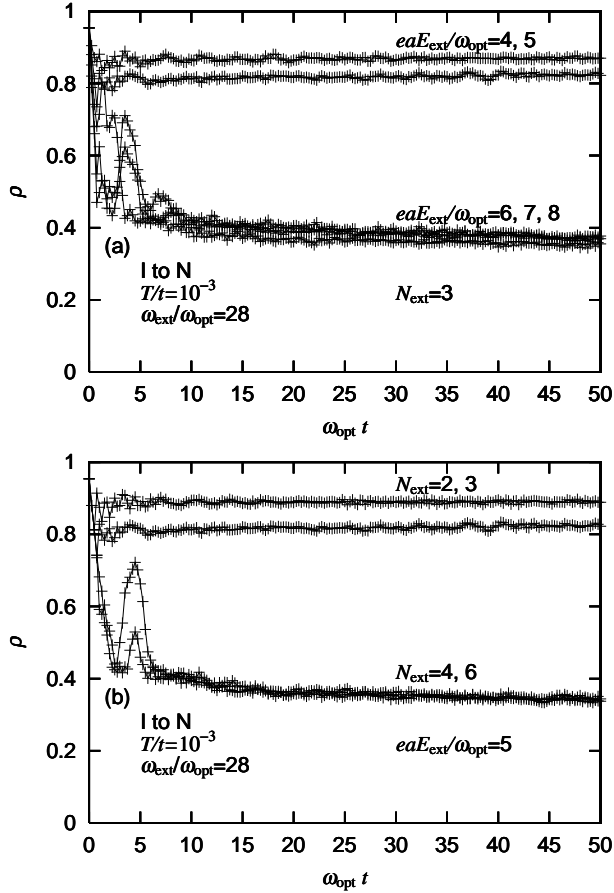}
\caption{Time dependence of the ionicity, (a) with different field strengths $ ( eaE_{\rm ext}/\omega_{\rm opt} )^2 $ for $ N_{\rm ext} $=3, and (b) with different pulse durations $ N_{\rm ext} $ for $ eaE_{\rm ext}/\omega_{\rm opt} $=5. The electric field of frequency $ \omega_{\rm ext} / \omega_{\rm opt} $=28 below the linear-absorption peak is applied [(a) for $ 0 \leq \omega_{\rm opt} t \leq N_{\rm ext} \omega_{\rm opt} T_{\rm ext} $=0.67, and (b) for $ 0 \leq \omega_{\rm opt} t \leq N_{\rm ext} \omega_{\rm opt} T_{\rm ext} $=0.45, 0.67, 0.90 and 1.35 with $ N_{\rm ext} $=2, 3, 4 and 6, respectively] to the ionic phase at $ T/t $=10$^{-3}$. The final state is either ionic or neutral.}
\label{fig:time_dependence}
\end{figure}

The evolution of the ionicity is shown in Fig.~\ref{fig:time_dependence}(a) with different field strengths $ ( eaE_{\rm ext}/\omega_{\rm opt} )^2 $, and in Fig.~\ref{fig:time_dependence}(b) with different pulse durations $ N_{\rm ext} $. Note that the oscillating electric field is applied in Fig.~\ref{fig:time_dependence}(a) for $ 0 \leq \omega_{\rm opt} t \leq N_{\rm ext} \omega_{\rm opt} T_{\rm ext} $=$ N_{\rm ext} \times 2 \pi /[ \omega_{\rm ext}/\omega_{\rm opt}] $=$ 3 \times 2 \pi/28 $=0.67. With weak fields below the threshold, the absorbed energy partially goes into the lattice oscillations, sometimes reversing the polarization of some ionic domains, and partially into the electronic part, increasing the deviation from the self-consistency condition, but the charge is not transferred so much between the donor molecules and the acceptor molecules. Above the threshold strength, a neutral domain is locally created before the field is turned off at $ t $=$ N_{\rm ext} T_{\rm ext} $. At this time, the spatially averaged ionicity is only slightly reduced if the field is just above the threshold. After that, the neutral domain grows, although the growth is not always monotonous due to the competition with surrounding ionic domains. In rare cases, as discussed later, an initially created neutral domain is completely annihilated and another neutral domain is created in a different place, which grows and finally covers the whole system. With stronger fields much above the threshold, the initially created neutral domain monotonously grows and considerably changes the spatially averaged ionicity already at $ t $=$ N_{\rm ext} T_{\rm ext} $. Similar evolution is observed by increasing the pulse duration. With short durations below the threshold, the ionicity is almost unchanged. Above the threshold duration, a neutral domain is created and grows after the field is turned off. In Fig.~\ref{fig:time_dependence}(b), the curves for $ N_{\rm ext} $=4 and 6 show this situation. In both cases, the neutral domain is transiently suppressed by surrounding ionic domains after $ t $=$ N_{\rm ext} T_{\rm ext} $. With a longer duration ($ N_{\rm ext} $=6), the system absorbs more energy and the transient suppression of the neutral domain is weaker. In any case, the threshold behavior shown in Figs.~\ref{fig:pulse_dependence_on_resonance} and \ref{fig:pulse_dependence_off_resonance} is a consequence of the fact that the ionicity either reaches a value in the neutral phase or almost returns to the initial value (i.e., no intermediate values between them) after the field is turned off, as clearly shown in Fig.~\ref{fig:time_dependence}.

\begin{figure}
\includegraphics[height=6cm]{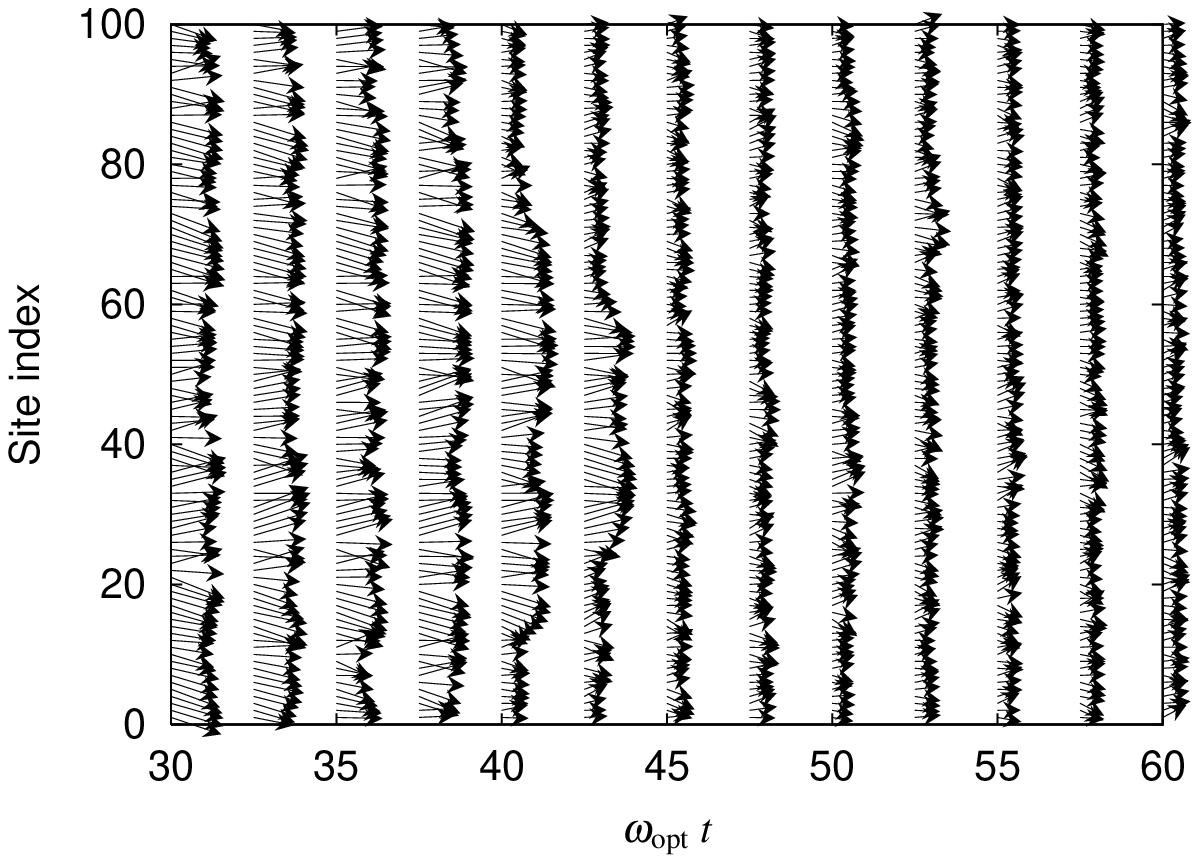}
\caption{Correlation between the staggered lattice displacement $ y_{st \; l} $ (the vertical component of the bar) and the ionicity $ \rho_l $ (the horizontal component of the bar), as a function of the site index $ l $ and the elapsing time $ t $ multiplied by $ \omega_{\rm opt} $. The electric field with $ eaE_{\rm ext}/\omega_{\rm opt} $=2, $ N_{\rm ext} $=150 and of frequency $ \omega_{\rm ext} / \omega_{\rm opt} $=28 below the linear-absorption peak is applied (for $ 0 \leq \omega_{\rm opt} t \leq N_{\rm ext} \omega_{\rm opt} T_{\rm ext} $=33.7) to the ionic phase at $ T/t $=10$^{-3}$. The final state is neutral.}
\label{fig:space_time_dependence}
\end{figure}

The space and time evolution of the ionicity and the staggered lattice displacement is shown in Fig.~\ref{fig:space_time_dependence}. The horizontal component of each bar represents the local ionicity $ \rho_l $ defined as $ \rho_l = 1 + (-1)^l ( - \langle n_{l-1} \rangle + 2 \langle n_l \rangle - \langle n_{l+1} \rangle )/4 $. The vertical component gives the local staggered lattice displacement $ y_{st \; l} $ defined as $ y_{st \; l} = (-1)^l ( - y_{l-1} + 2 y_l - y_{l+1} )/4 $. The bars are shown on all sites and selected times. Now the oscillating electric field is applied for $ 0 \leq \omega_{\rm opt} t \leq 33.7 $. The initial ionic phase at $ t $=0 has uniformly downward bars (not shown). Note that, for $ \omega_{\rm opt} t \leq 33.7 $, some ionic domains have reversed polarizations. When the field is turned off, the local ionicity is weakened around $ 90 < l <100 $. This local area can be regarded as the nucleus or the seed of a neutral domain, although the ionicity of this area at $ \omega_{\rm opt} t = 33.7 $ is still larger than in the neutral domain appearing later. Once the neutral domain appears here, it quickly grows and covers the 100-site system within one or two periods of the optical lattice vibration (whose period is about $ 4 / \omega_{\rm opt} $). The boundary between the ionic and neutral domains is clearly seen because the ionicity changes discontinuously at the boundary. Immediately after the transition is completed, the ionicity is low but still spatially fluctuates. With increasing $ \omega_{\rm opt} t $, the ionicity becomes spatially uniform.

In order to analyze the time-dependent electronic state, we obtained Fourier transforms of the ionicity in ref.~\citen{miyashita03} to compare with them the current excitation spectra of the static neutral state in the RPA. The RPA is equivalent to the time-dependent HF approximation used here but limited to infinitesimal deviations from the static HF solution, so that the linear absorption peak in Fig.~\ref{fig:linear_absorption} corresponds to the peak in the RPA spectrum due to delocalized excitons. After a sufficiently large number of electrons are excited, the time-dependent ionicity has high-frequency components due to delocalized excitons and low-frequency components due to complex motion of the neutral-ionic domain walls \cite{miyashita03}. When the number of excited electrons decreases, the correspondence to delocalized excitons becomes worse because excitons are more sensitively affected by lattice fluctuations \cite{miyashita_D}. Since the frequency of the electric field applied in Fig.~\ref{fig:space_time_dependence} is below the peak in Fig.~\ref{fig:linear_absorption}, the photoirradiation produces excitons localized by lattice fluctuations.

Although the time-dependent Schr\"odinger equation governs the deterministic time-evolution, the positions where excitons are produced are distributed in a stochastic manner according to the random numbers initially added to the lattice variables. As seeds of stable domains generally trigger a first-order transition, localized excitons promote the transition. Indeed, the localized excitons more efficiently nucleate neutral domains and trigger the transition (Fig.~\ref{fig:pulse_dependence_off_resonance}) than the delocalized excitons (Fig.~\ref{fig:pulse_dependence_on_resonance}). Neutral domains are created by the localized excitons, so that they are also distributed at random. Although these nucleation processes are stochastic, the neutral domains grow in a deterministic manner as long as the neutral-ionic domain walls do not collide with ionic-anti-ionic (I-\=I) solitons, which are boundaries between different polarizations inside the ionic domain, as discussed in ref.~\citen{miyashita03}. That is why the coherence can appear, as discussed later in paper III. Namely, if the stochastic processes dominated the time-evolution of the system, the coherence would be lost.

\subsection{Strong vs. weak pulses}\label{subsec:strong_vs_weak}

\begin{figure}
\includegraphics[height=12cm]{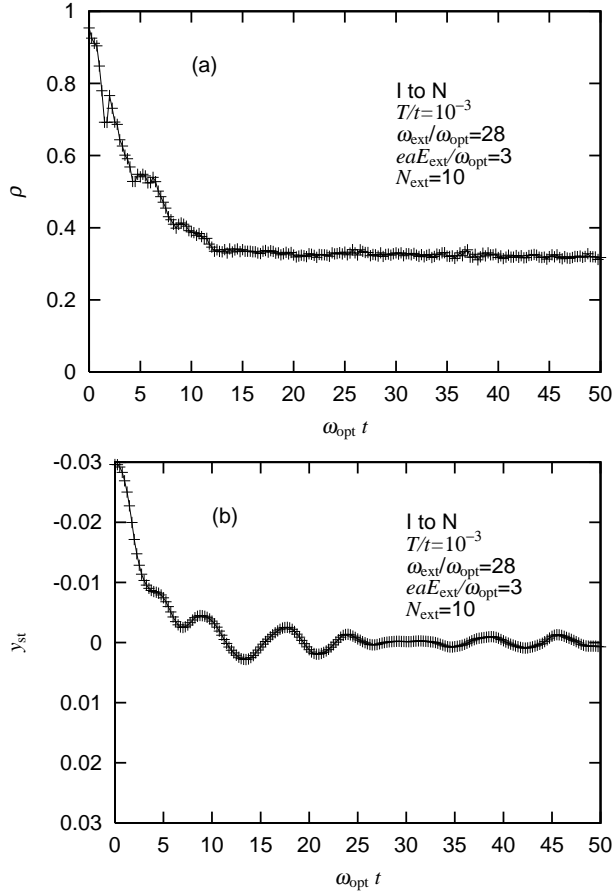}
\caption{Time dependence of (a) the ionicity and (b) the staggered lattice displacement when the pulse of frequency $ \omega_{\rm ext} / \omega_{\rm opt} $=28 is rather strong, $ eaE_{\rm ext}/\omega_{\rm opt} $=3, and short, $ N_{\rm ext} $=10 (i.e., $ 0 \leq \omega_{\rm opt} t \leq N_{\rm ext} \omega_{\rm opt} T_{\rm ext} $=2.24). The initial state is ionic at $ T/t $=10$^{-3}$. The ionicity and the staggered lattice displacement change on the same time scale to reach the neutral phase.}
\label{fig:1_step_transition}
\end{figure}

To see how the electronic dynamics and the lattice dynamics are correlated and how their correlation can be controlled by changing the parameters of the oscillating electric field, we use different strengths and durations of pulses to calculate the time evolution of the ionicity and of the staggered lattice displacement. With a strong and short pulse, the electronic dynamics and the lattice dynamics are closely correlated. One example is shown in Fig.~\ref{fig:1_step_transition}. On a magnified scale, one sees small-amplitude and very rapid modulation of the ionicity [Fig.~\ref{fig:1_step_transition}(a)] related with excitonic excitations \cite{miyashita03}, in contrast to the slow oscillation of the staggered lattice displacement [Fig.~\ref{fig:1_step_transition}(b)]. However, the overall evolutions of the ionicity and of the staggered lattice displacement are similar to each other.

\begin{figure}
\includegraphics[height=12cm]{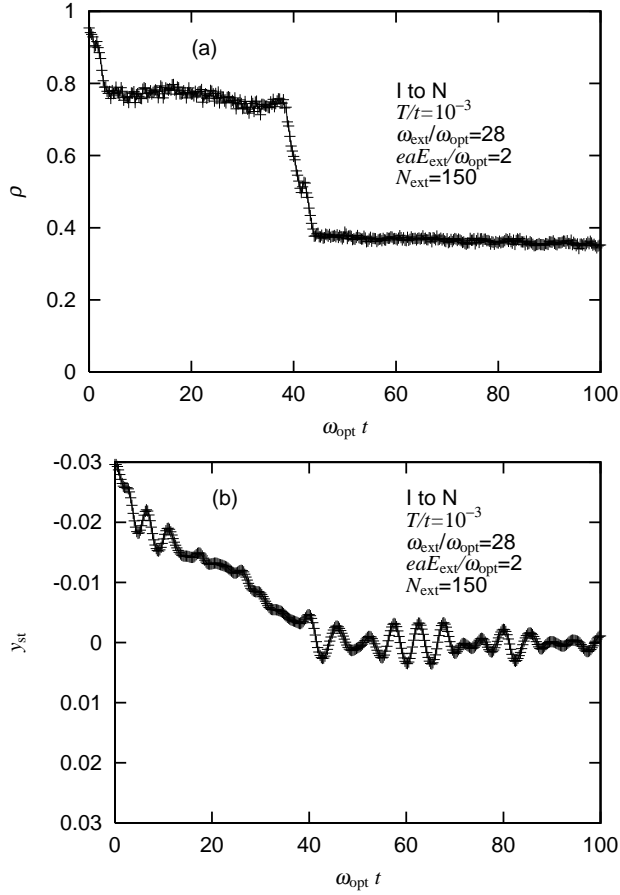}
\caption{Time dependence of (a) the ionicity and (b) the staggered lattice displacement when the pulse of frequency $ \omega_{\rm ext} / \omega_{\rm opt} $=28 is weaker, $ eaE_{\rm ext}/\omega_{\rm opt} $=2, and longer, $ N_{\rm ext} $=150 (i.e., $ 0 \leq \omega_{\rm opt} t \leq N_{\rm ext} \omega_{\rm opt} T_{\rm ext} $=33.7), than in Fig.~\ref{fig:1_step_transition}. The initial state is ionic at $ T/t $=10$^{-3}$. The staggered lattice displacement decays faster than the ionicity to reach the neutral phase.}
\label{fig:1and1/2_step_transition}
\end{figure}

As the pulse becomes weaker and longer, their evolutions become different from each other, as shown in Fig.~\ref{fig:1and1/2_step_transition}. The ionicity initially drops a little bit, but it remains large for a while [Fig.~\ref{fig:1and1/2_step_transition}(a)]. During the same period, the staggered lattice displacement steadily decreases and almost vanishes [Fig.~\ref{fig:1and1/2_step_transition}(b)]: the polarization of some ionic domain is reversed, then that of another domain is reversed, and so on. In other words, I-\=I solitons separating ionic domains with different polarizations are created in a random manner during this period. When the staggered lattice displacement vanishes on average, the system is regarded as in the paraelectric ionic phase. It should be noted that this phase is known to appear in TTF-CA only under high pressure \cite{MHLemee-C97} in thermal equilibrium. The mechanism of its appearance is theoretically clarified in the context of the electrostriction effect \cite{kishine04}. Furthermore, after intra-molecular photoexcitations, this phase is suggested to appear by comparing the photoreflectance and the second-harmonic-generation signal \cite{luty02} and by the x-ray diffraction \cite{guerin04}. The present numerical result suggests that, even by intra-chain charge-transfer excitations, it is possible to transiently create the paraelectric ionic phase if the pulse is so weak and long that I-\=I solitons are created before the charge transfer really occurs. When the pulse is weak, the supplied energy is not used to directly transfer charge density along the chain. By intramolecular excitations, the supplied energy would not be directly transferred to intra-chain charge transfer processes, either. In this sense, the paraelectric ionic phase may appear by a common mechanism. If the electric field continues to be applied, this phase is finally converted into the neutral phase.

\begin{figure}
\includegraphics[height=12cm]{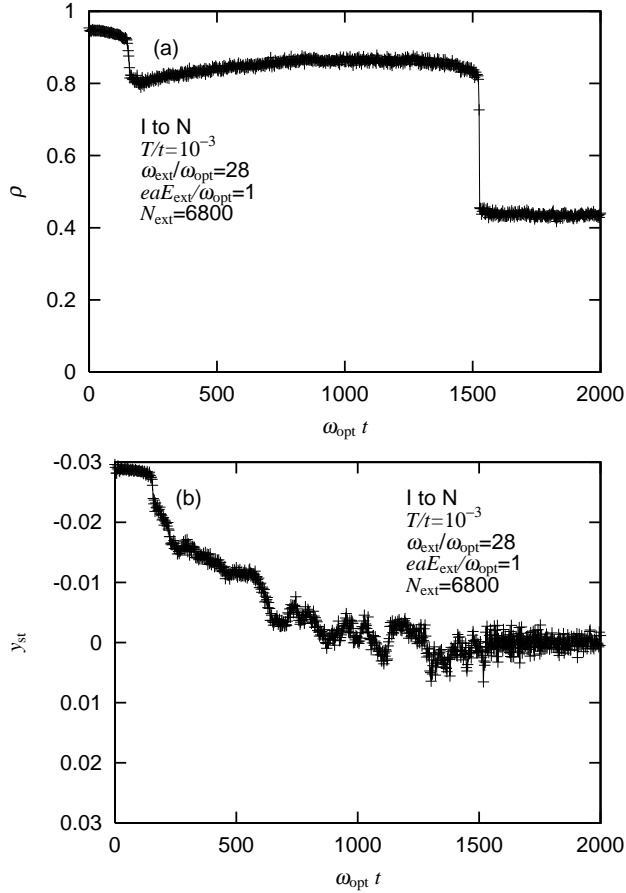}
\caption{Time dependence of (a) the ionicity and (b) the staggered lattice displacement when the pulse of frequency $ \omega_{\rm ext} / \omega_{\rm opt} $=28 is much weaker, $ eaE_{\rm ext}/\omega_{\rm opt} $=1, and much longer, $ N_{\rm ext} $=6800 (i.e., $ 0 \leq \omega_{\rm opt} t \leq N_{\rm ext} \omega_{\rm opt} T_{\rm ext} $=1530), than in Fig.~\ref{fig:1_step_transition}. The initial state is ionic at $ T/t $=10$^{-3}$. The ionic phase without long-ranged dimerization pattern (thus paraelectric) lasts long before the neutral phase is reached.}
\label{fig:2_step_transition}
\end{figure}

The weak-pulse case where the supplied energy dissipates more into I-\=I solitons is similar to the large-$ T $ case in ref.~\citen{miyashita03}. When the initial lattice fluctuations are small, few I-\=I solitons are present. The neutral domain grows smoothly and rapidly (Fig.~5 of ref.~\citen{miyashita03}). Otherwise many I-\=I solitons collide with the neutral-ionic domain walls to make the motion of the latter irregular. Then, the growth of the neutral domain becomes rather chaotic (Fig.~7 of ref.~\citen{miyashita03}). In this paper, many I-\=I solitons are created when the weak field is applied for a long time, and the phases of the staggered lattice displacements (i.e., the polarizations) are disordered to average out the broken inversion symmetry. Accordingly, the growth of the neutral domain becomes rather irregular, e.g., in the case of Fig.~\ref{fig:2_step_transition}, although it is invisible on this scale.

When the very weak electric field is applied continuously, the ferroelectric ionic phase survives for a while, as shown in Fig.~\ref{fig:2_step_transition}. Then, the ionicity drops a little bit, but further charge transfer does not take place for a long time [Fig.~\ref{fig:2_step_transition}(a)], while the absorbed energy slowly disorders the polarization of the ionic phase to reduce and finally destroy the long-range order of the staggered lattice displacement [Fig.~\ref{fig:2_step_transition}(b)]. It takes a long time to suppress the staggered lattice displacement and an even longer time to create a neutral domain. The paraelectric ionic phase has a long lifetime. It can be made infinite within the computational time scales if the field is turned off before the neutral domain is created. Once the neutral domain is created, it grows as a whole. The I-\=I solitons obstruct the growth of the neutral domain \cite{miyashita03}, but they cannot completely stop it.

\begin{figure}
\includegraphics[height=12cm]{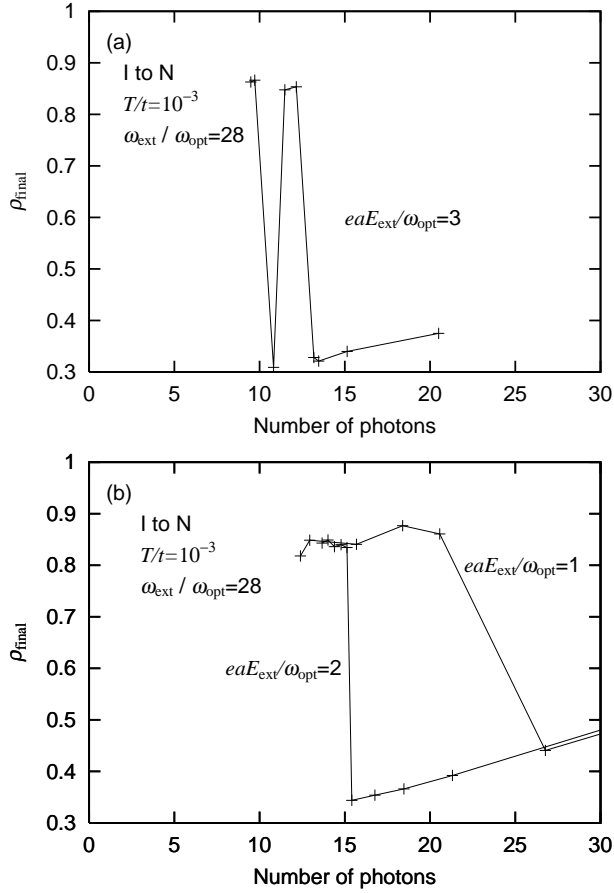}
\caption{Final ionicity as a function of the number of absorbed photons, (a) for pulses of intermediate strength, $ eaE_{\rm ext}/\omega_{\rm opt} $=3, and (b) for weak pulses, $ eaE_{\rm ext}/\omega_{\rm opt} $=1 and 2. The electric field of frequency $ \omega_{\rm ext} / \omega_{\rm opt} $=28 is applied to the ionic phase at $ T/t $=10$^{-3}$.}
\label{fig:strength_dependent_threshold}
\end{figure}

As stated above, when the weak field is applied for long, the supplied energy dissipates more into I-\=I solitons and others, so that it is not used to directly transfer charge density along the chain. It somewhat modifies the threshold relation. The final ionicity is then plotted in Fig.~\ref{fig:strength_dependent_threshold}, as a function of the number of absorbed photons, for intermediate to weak pulses. The threshold absorption is indeed shifted to larger values. Namely, it is about 11 photons per 100 sites for $ eaE_{\rm ext}/\omega_{\rm opt} $=5 [Fig.~\ref{fig:pulse_dependence_off_resonance}(d)], 11 or 13 photons for $ eaE_{\rm ext}/\omega_{\rm opt} $=3 [Fig.~\ref{fig:strength_dependent_threshold}(a)], 15 photons for $ eaE_{\rm ext}/\omega_{\rm opt} $=2, and 27 photons for $ eaE_{\rm ext}/\omega_{\rm opt} $=1 [Fig.~\ref{fig:strength_dependent_threshold}(b)]. In Fig.~\ref{fig:strength_dependent_threshold}(a), the final ionicity is not a monotonous function. As the pulse duration increases, it once drops and returns to a large value before it finally drops again to a small value. This is because the initially created neutral domain is annihilated by continuing photoirradiation. It never happens if the field is strong. For intermediate to weak pulses, larger dissipation or larger excess energy can disturb the growth of the neutral domain. The second neutral domain, which finally grows and covers the whole system, is created in a different place from the first one.

Here we estimate the energy flow again. It is true that many I-\=I solitons are produced for weak and long pulses, but their creation energies are much smaller than those of the neutral-ionic domain walls mentioned before. Both of the lattice kinetic and elastic energies are finally in the range between 0.1eV and 0.2eV for the 100-site chain. It implies that most of the supplied energy is absorbed into the electronic state. Consequently, the final ionicity deviates from 0.20 for the static self-consistent solution in the metastable neutral phase: it is 0.32 for $ eaE_{\rm ext}/\omega_{\rm opt} $=3 (Fig.~\ref{fig:1_step_transition}), 0.35 for $ eaE_{\rm ext}/\omega_{\rm opt} $=2 (Fig.~\ref{fig:1and1/2_step_transition}), and 0.43 for $ eaE_{\rm ext}/\omega_{\rm opt} $=1 (Fig.~\ref{fig:2_step_transition}).

\subsection{Infrared pulses}\label{subsec:infrared_pulses}

Because of the electron-lattice coupling, photons whose frequencies are about that of the optical lattice vibration are also absorbed and can contribute to the phase transition. The transition to the neutral phase is induced by sufficiently strong irradiation of the infrared light. By setting the frequency of the electric field $ \omega_{\rm ext} $ comparable with that of the optical lattice vibration, we vary the field strength $ ( eaE_{\rm ext} )^2 $ and the pulse duration $ N_{\rm ext} T_{\rm ext} $. Below threshold strength or duration, photons are hardly absorbed. Above the threshold values, many photons are absorbed. The important point here is again the fact that, once substantial absorption takes place, the ionicity substantially decreases. Besides the fact that the relation between the number of absorbed photons and either the field strength or the pulse duration is now strongly nonlinear (not shown), the presence of the threshold and the behavior above the threshold are very similar to those in the case of excitonic excitations.

\section{Conclusions}\label{conc}

In the quasi-one-dimensional mixed-stack organic charge-transfer complex, TTF-CA, photoirradiation is known to trigger transitions from the ionic to neutral phases and from the neutral to ionic phases. In the former transition, the coherent motion of the macroscopic neutral-ionic domain boundary is observed as well as the threshold density of absorbed photons \cite{iwai02}. Coherence will be discussed later. To deterministically approach the dynamics of charge density coupled with lattice displacements during this transition, we employ the one-dimensional extended Peierls-Hubbard model with alternating potentials and numerically solve the time-dependent Schr\"odinger equation for the electronic wave function at the mean-fled level combined with the classical equation of motion for the lattice displacements. To treat the photoexcitations more realistically than in the previous study \cite{miyashita03}, we incorporate a pulse of oscillating electric field into the Peierls phase of the transfer integral and vary the frequency, the amplitude, and the duration of the pulse independently. 

When the dimerized ionic phase is photoexcited, we reproduce the threshold behavior in the final ionicity as a function of the increment of the total energy, which is regarded as proportional to the density of absorbed photons. The threshold photon density rather weakly depends on the amplitude and the duration of the pulse. However, when the final ionicity is plotted as a function of the amplitude or duration of the pulse, we found that the pulse with frequency below the linear absorption peak is more effective than that at the peak. The phase transition proceeds by nucleation of a metastable neutral domain and its rapid growth. As a consequence, after the electric field is turned off, the electronic state either reaches the neutral one (above the threshold photoexcitation) or returns to the ionic one with excess energy supplied by photoexcitations (below the threshold), but basically no intermediate state is produced. The efficiency of the pulse with frequency below the linear absorption peak originates from the fact that lattice-fluctuation-induced localized excitons are more effective on nucleating the neutral domain than delocalized excitons. 

Here we discuss coherence. Below and in the following papers, we use this word in the numerical context to describe a state where the phases of the staggered lattice displacements are almost uniform in a wide spatial and temporal region inside the ionic domain, while the molecules are essentially equidistant in the neutral domain, so that the space and time dependent ionicity matches the staggered displacements averaged over the corresponding region. When the phases of the staggered displacements are almost uniform in the ionic background, few I-\=I solitons are present. Thus the neutral domain grows smoothly and rapidly (Fig.~5 of ref.~\citen{miyashita03}). Otherwise many I-\=I solitons are present, colliding with the neutral-ionic domain walls. Then, the growth of the neutral domain is rather irregular (Fig.~7 of ref.~\citen{miyashita03}). In these examples, the coherence is strong in the former, and weak in the latter.

In this paper, the coherence is shown to be controlled by the field strength and the pulse duration. When the pulse is strong and short, the ionic-to-neutral transition is achieved without producing many I-\=I solitons. The charge transfer from the acceptor to donor molecules takes place on the same time scale with the disappearance of dimerization. When the pulse is weak and long, however, many I-\=I solitons are produced. The dimerization-induced polarization is disordered and the staggered lattice displacements disappear on average to restore the inversion symmetry before the charge transfer takes place to bring the system neutral. Thus, the intermediate phase is regarded as paraelectric and ionic. The lifetime of the paraelectric ionic phase becomes long as the pulse becomes weakened. In paper III, the coherence is shown to be indeed strong when the pulse is strong and short. The phases of the staggered lattice displacements remain to be almost uniform long after the photoirradiation, so that an interference effect is caused by a double pulse.

In addition to the excitonic excitations summarized above, we apply pulses with infrared frequencies to demonstrate that the transition is induced also when the pulse resonates with the optical lattice vibrations. The threshold behavior is again observed. In paper III, the interference is shown to be closely related with the optical lattice vibrations.

Let us come back to the coherence observed experimentally \cite{iwai02}. The used pulse would correspond to the strong and short pulse in the theoretical context. Then, the mainly ionic state before the transition would have very low density of I-\=I solitons, some of which are already present by thermal fluctuations before the photoirradiation. Thus three-dimensional ferroelectric order initially survives for a while. After a significant density of neutral domains are created by the photoirradiation, if the density of I-\=I solitons would remain quite low during the proliferation of neutral domains, the phases of the neutral-ionic domain walls can in principle be matched (i.e., their motion is coherent), allowing to be experimentally observed as oscillating photoreflectance. If high density of I-\=I solitons were produced after the photoirradiation, on the other hand, the motion of the neutral-ionic domain walls would be irregular, making the experimental observation hard. These arguments are of course based on the numerically derived assumption that the motion of the neutral-ionic domain walls is sensitively affected by collisions with I-\=I solitons.

We compare the motion of the neutral-ionic domain walls with the so-called domino effect. A spontaneous growth after some trigger is often described as the domino effect, but it needs caution. Propagation of structural changes is theoretically realized in a one-dimensional lattice model coupled with localized electrons after an electron is excited if the strength and range of the elastic coupling are appropriate \cite{koshino98}. It proceeds site by site, i.e., a displacement at a certain site causes another displacement at the next site, then at its next site, and so on. At each site, the lattice is displaced under friction on the adiabatic potential. In the ionic-to-neutral transition, the one-dimensional dynamics is similar to this domino-like behavior only in the sense that, once triggered properly, a new domain spontaneously grows accompanying the structural changes. However, the friction is so weak that the phases of the staggered lattice displacements are indeed important in the transition dynamics, as shown in paper III in the context of the interference effect. Thus, the coherent dynamics of the present model is very different from the domino-like propagation of structural changes in the model of ref.~\citen{koshino98}. Another difference is in the itineracy of electrons in the present model, where both electrons and lattice displacements show cooperative and coupled dynamics.

When comparing the simulated dynamics with the experimental findings in more detail, we would need to take account of three-dimensionality and energy dissipation. If inter-chain couplings are significant, small neutral domains in the ionic background are energetically unfavorable. Then sufficiently large density of neutral domains would be needed to suppress the effect of inter-chain couplings and to further increase the density of neutral domains. In the one-dimensional model used in this paper, once metastable neutral domains appear, they easily grow without energy loss from surroundings if the neutral and ionic phases are almost degenerate in equilibrium. Therefore, we cannot quantitatively compare the threshold behavior. Nevertheless, neutral domains grow first along the chains, so that the present study would be an important step toward more detailed descriptions. 

If energy dissipation is not negligible, a part of the energy supplied by the photoexcitation flows into electrons in different orbitals or different lattice vibrations (including molecular ones) from those which are considered in the present model. As the transition takes more time (i.e., as the pulse becomes weaker), the effect of the dissipation would be larger and negative as long as the ionic-to-neutral transition is concerned. Indeed, the relation between the strength and the duration of the pulse at the threshold in the present isolated system deviates for weak and long pulses from that derived for a classical model with the help of a stochastic equation \cite{nagaosa89}. It implies that the supplied energy is effectively used by prohibiting the energy flow into surroundings. 

In general, these effects of three-dimensionality and energy dissipation would become substantial for the long-time behavior. However, the recently observed short-time behavior would not suffer from them so much. The present result concerning the strongly nonlinear relation between the final ionicity and the density of absorbed photons would be qualitatively unchanged, although three-dimensionality or energy dissipation may quantitatively modify the relation. It would be worth experimental investigation to change the field strength to seek a possibility for changing the ionicity and staggered lattice displacements simultaneously or separately in a controlled manner.

\section*{Acknowledgement}

The author is grateful to S. Koshihara, T. Luty and H. Okamoto for showing their data prior to publication and for enlightening discussions. He appreciates the assistance of N. Miyashita at the early stage of the present work. 
This work was supported by Grants-in-Aid for Scientific Research (C) (No. 15540354), for Scientific Research on Priority Area ``Molecular Conductors'' (No. 15073224), for Creative Scientific Research (No. 15GS0216), and NAREGI Nanoscience Project from the Ministry of Education, Culture, Sports, Science and Technology, Japan.


\end{document}